\documentclass{ws-ijmpa}

\begin{document}

\newcommand\be{\begin{equation}}
\newcommand\ee{\end{equation}}
\newcommand\bea{\begin{eqnarray}}
\newcommand\eea{\end{eqnarray}}
\newcommand\bseq{\begin{subequations}} 
\newcommand\eseq{\end{subequations}}
\newcommand\bcas{\begin{cases}}
\newcommand\ecas{\end{cases}}
\newcommand{\p}{\partial}
\newcommand{\f}{\frac}


%
\catchline{}{}{}{}{}
%

\title{Quantum cosmology with a minimal length}

\author{Marco Valerio Battisti}

\address{ICRA - International Center for Relativistic Astrophysics\\Phys. Dept. (G9), University of Rome ``La Sapienza'' P.le A. Moro 5, 00185 Rome, Italy\\
battisti@icra.it}

\author{Giovanni Montani}

\address{ICRA - International Center for Relativistic Astrophysics\\Phys. Dept. (G9), University of Rome ``La Sapienza'' P.le A. Moro 5, 00185 Rome, Italy\\ENEA C.R. Frascati (Dipartimento F.P.N.), Via Enrico Fermi 45, 00044 Frascati, Rome, Italy\\ICRANET C.C. Pescara, P.le della Repubblica 10, 65100 Pescara, Italy\\
montani@icra.it}

\maketitle

\begin{abstract}
Quantum cosmology in the presence of a fundamental minimal length is analyzed in the context of the flat isotropic and the Taub cosmological models. Such minimal scale comes out from a generalized uncertainty principle and the quantization is performed in the minisuperspace representation. Both the quantum Universes are singularity-free and (i) in the isotropic model no evidences for a Big-Bounce appear; (ii) in the Taub one a quasi-isotropic configuration for the Universe is predicted by the model.

\end{abstract}

\ccode{PACS numbers: 98.80.Qc, 11.10.Nx, 04.20.Dw}

\bigskip

The existence of a fundamental cut-off length has long been expected in quantum gravity and recently it has been proposed how it can appear by modifying the Heisenberg uncertainty relation by the so-called generalized uncertainty principle (GUP) [\refcite{Kem}]
\be\label{gup}
\Delta q \Delta p\geq \f 1 2\left(1+\beta (\Delta p)^2+\beta \langle{\bf p}\rangle^2\right),
\ee
where $\beta$ is a ``deformation'' parameter. As matter of fact, it is immediate to verify that such a relation (\ref{gup}) implies a finite minimal uncertainty in position $\Delta q_{min}=\sqrt\beta$. This way, we claim that this approach entail a minimal scale in the quantum framework. However, the cut-off predicted by the GUP is, by its nature, different from the minimal length predicted by other approaches, for example the minimal eigenvalue of the geometric operators in loop quantum gravity [\refcite{Thi}]. 

The relation (\ref{gup}) has been appeared in the context of string theory, where a minimal observable length it is a consequence of the fact that strings can not probe distance below the string scale [\refcite{String}]. However, recently, a wide work has been made on this field in a large variety of directions (see for example [\refcite{GUP1}] and the references therein).

This paper is devoted to review some results obtained in a recent approach to quantum cosmology, in which the GUP framework was applied to the Universe minisuperspace dynamics [\refcite{BM07c}]. In particular, in our previous works the flat isotropic cosmological model [\refcite{BM07a}] and the Taub one [\refcite{BM07b}] are been analyzed in such a scheme. The application of this framework in quantum cosmology appear to be physically well grounded. In fact, the generalized uncertainty principle (\ref{gup}) can be immediately reproduced modifying the canonical Heisenberg algebra by the following one
\be\label{modal}
[{\bf q},{\bf p}]=i(1+\beta{\bf {p}}^2). 
\ee 
Although such a deformed commutation relation, differently from the GUP itself, has not been so far derived directly from string theory, it is a possible way in which certain features of a more fundamental theory may manifest themselves in some toy models (finite degrees of freedom). As well known, by the minisuperspace representation the phase space of General Relativity (GR) is truncated and a field theory is reduced to a mechanical system. In particular, the homogeneous sector of GR, i.e. the Bianchi models, are characterized by three degrees of freedom (the three scale factors) and the isotropic one, i.e. the FRW models, by a single one. In this respect, as the GUP approach relies on a modification of the canonical quantization prescription, it can be reliably applied to any dynamical system, i.e. also to the cosmological models. Furthermore, by such a formalism, some features of string theory will be implemented into the early Universe dynamics. 

The appearance of a nonzero uncertainty in position pose some difficulty in the construction of an Hilbert space. In fact, as well-known, no physical state which is a position eigenstate can be constructed since an eigenstate of an observable has necessarily to have vanishing uncertainty on it. Although it is possible to construct position eigenvectors, they were only formal eigenvectors but not physical states. To be more precise, let us assume the commutation relations to be represented on some dense domain $D\subset\mathcal H$ in a Hilbert space $\mathcal H$. In the canonical case, a sequence $\vert\psi_n\rangle\in D$ with position uncertainties decreasing to zero, exists. On the other hand, in presence of a minimal uncertainty $\Delta q_{min}>0$, it is not possible any more to find some $\vert\psi_n\rangle\in D$ such that
\be
\lim_{n\rightarrow\infty}\left(\Delta q_{min}\right)_{\vert\psi_n\rangle}=\lim_{n\rightarrow\infty}\langle\psi\vert({\bf q}-\langle\psi\vert{\bf q}\vert\psi\rangle)^2\vert\psi\rangle=0.
\ee  
Therefore, at the GUP quantum level {\it no physical states which are position eigenstates exist at all} and so we lost direct information on the position itself. The knowledge on it can be recovered only by the study of the states which realize the maximally-allowed localization. The Heisenberg algebra (\ref{modal}) can be represented in the momentum space, where the $\bf q$, $\bf p$ operators act as 
\be\label{rep}
{\bf p}\psi(p)=p\psi(p), \qquad {\bf q}\psi(p)=i(1+\beta p^2)\p_p\psi(p),
\ee
on a dense domain $S$ of smooth functions. The maximal localization states $\vert\psi^{ml}_{\zeta}\rangle$ can be constructed from the minimal uncertainty relation $\Delta q\Delta p=\f12\vert\langle [{\bf q},{\bf p}]\rangle\vert$ and they satisfy the proprieties: $\langle\psi^{ml}_{\zeta}\vert {\bf q}\vert\psi^{ml}_{\zeta}\rangle=\zeta$ and $(\Delta q)_{\vert\psi^{ml}_{\zeta}\rangle}=\Delta q_{min}$. These states are proper physical ones around the position $\zeta$, the so-called quasiposition. To obtain the probability amplitude to find a particle maximally localized around $\zeta$, i.e. with the standard deviation $\sqrt\beta$, we have to project an arbitrary state $\vert\psi\rangle$ on $\vert\psi^{ml}_{\zeta}\rangle$ and derive the {\it quasiposition wave function} $\psi(\zeta)\equiv\langle\psi^{ml}_{\zeta}\vert\psi\rangle$
\be\label{qwf} 
\psi(\zeta)\sim\int\f{dp}{(1+\beta p^2)^{3/2}}e^{i\f{\zeta}{\sqrt{\beta}} \tan^{-1}(\sqrt{\beta}p)}\psi(p).
\ee
This is nothing but a generalized Fourier transformation, where in the $\beta=0$ limit the ordinary position wave function $\psi(\zeta) = \langle\zeta\vert\psi\rangle$ is recovered. For details on this construction see [\refcite{Kem}].

As we said this framework is used to quantize a cosmological model and to compare it with the Wheeler-DeWitt (WDW) approach. In particular we are interested on the fate of the cosmological singularity at quantum level, since it is no tamed by the (quantum) canonical effects. Firstly we have to clarify a general criteria for determining whether the quantized models actually collapse [\refcite{Got}]. This is a non trivial question and there is not such a rigorous criteria yet. As well-known a space-time singularity in GR is defined by two criteria [\refcite{HE}]. The first one is the causal geodesic incompleteness (global criteria) and the second one is the divergence of the scalars built up to the Riemann tensor (local criteria). Although the latter one is useful to characterize a singularity, it is unsatisfactory since a space-time can be singular without any pathological character of these scalars. At quantum level the task is more difficult. The early idea by DeWitt was to impose the condition that the wave function vanishes at the singularity [\refcite{Got}], but this boundary conditions has little to do with the quantum singularity avoidance. The accepted criteria for a singularity-free model is the non-fall of the wave packets in the classical singularity. This way, the probability to find the Universe in this non-physical region is negligible.

Let us now analyze the flat FRW model coupled with a massless scalar field $\phi$ [\refcite{BM07a}]. The dynamics of this model is summarized in the scalar constraint
\be\label{con}
H\equiv H_{grav}+H_{\phi}=-9\kappa p_x^2x+\f3 {8\pi}\f{p_{\phi}^2}{x}\approx0 \quad x\equiv a^3,
\ee
where $\kappa=8\pi G\equiv8\pi l_P^2$ is the Einstein constant and $a$ is the scale factor. The phase space is $4$-dimensional, with coordinates $(x,p_x;\phi,p_{\phi})$ and at $x=0$ the physical volume of the Universe goes to zero and the singularity appears. Since $\phi$ does not enter the expression of the constraint, $p_{\phi}$ is a constant of motion and therefore each classical trajectory can be specified in the $(x,\phi)$-plane. Thus $\phi$ can be considered as a relational time and the dynamical trajectory reads as
\be\label{clastra}
\phi=\pm\f 1 {\sqrt{24\pi\kappa}}\ln\left|\f x {x_0}\right|+\phi_0,
\ee
where $x_0$ and $\phi_0$ are integration constants. In this equation, the plus sign describes an expanding Universe from the Big-Bang, while the minus sign a contracting one into the Big-Crunch. We stress that the classical cosmological singularity is reached at $\phi=\pm\infty$ and every classical solution gets to it. At quantum level the WDW equation $(\p^2_\phi+\widehat\Theta)\Psi=0$ ($\Theta\equiv24\pi\kappa(xp_x^2x)$), coming out from the constraint (\ref{con}), tells us how the wave function $\Psi=\Psi(x,\phi)$ evolves as $\phi$ changes, i.e. we can regard the argument $\phi$ of $\Psi(x,\phi)$ as an ``emergent time'' and the scale factor as the real physical variable. In order to have an explicit Hilbert space, we perform the natural decomposition of the solution into positive and negative frequency parts. Therefore, the solution of the WDW equation has the very well-known form
\be\label{solcan} 
\Psi_\epsilon(x,\phi)=x^{-1/2}\left(Ax^{-i\gamma}+Bx^{i\gamma}\right)e^{i \sqrt{24\pi\kappa}\epsilon\phi}, 
\ee
where $\gamma=\f 12(4\epsilon^2-1)^{1/2}\geq0$ and $\epsilon^2$ being the eigenvalue of the operator $\hat\Theta$. Thus the spectrum is purely continuous and covers the interval $(\sqrt{3}/2l_P,\infty)$. The wave function $\Psi_\epsilon(x,\phi)$ is of positive frequency with respect to the internal time $\phi$ and it satisfies the square root of the quantum constraint (\ref{con}): we deal with a Sch\"odinger-like equation $i\p_\phi\Psi=-\sqrt{\hat\Theta}\Psi$. It is no difficult to note, that such an approach to this problem does not solve the singularity problem. More precisely, from the solution (\ref{solcan}) it is possible to construct a localized state at some initial time and, in the backward evolution toward the Big-Bang, its peak will move along the classical trajectory (\ref{clastra}), i.e. it will fall into the classical singularity. This way, the Big-Bang singularity is not tamed by the WDW framework.

Also in the GUP framework, we regard the scalar field as an ``emergent time'' for the quantum evolution and then we treat in the ``generalized'' way only the real degree of freedom of the problem: the isotropic volume $x$. Therefore, the couple of conjugate variables ($\phi,p_\phi$) is canonically quantized and the deformed WDW equation is $(\p^2_\phi+\widehat\Theta_{gup})\Psi=0$, where the action of $\widehat\Theta_{gup}$ on $\Psi(p)$ is
\be\label{emu}
\mu^2(1+\mu^2)^2\f{d^2\Psi}{d\mu^2}+2\mu(1+\mu^2)(1+2\mu^2)\f{d\Psi}{d\mu}+\epsilon^2\Psi=0,
\ee  
$\mu\equiv\sqrt\beta p$ being a dimensionless parameter. As before we have decomposed the solution of the deformed WDW equation into positive and negative frequency parts and focus on the positive frequency sector, i.e. we have considered $\overline{\Psi}(p,\phi)=\Psi(p)e^{i \sqrt{24\pi\kappa}\epsilon\phi}$, where $p\equiv p_x$. The equation (\ref{emu}) is solved by two changes of variables, $\rho\equiv\tan^{-1}\mu$ ($\mu\in[0,\infty)\Rightarrow\rho\in[0,\pi/2]$) and $\xi\equiv\ln(\sin\rho)$ ($\xi\in(-\infty,0]$), and the solution reads 
\be
\Psi_\epsilon(\xi)=C e^{-\xi(1-\alpha)}\left(1+be^{2\xi}\right),
\ee
where $\alpha=\sqrt{1-\epsilon^2}$ and $b=(1-\alpha)/(1+\alpha)$. The {\it quasiposition wave function} (\ref{qwf}) relative to this problem is 
\be\label{quapos}
\Psi_\epsilon(\zeta)=\int_{-\infty}^0d\xi \exp{\left(\xi+i\zeta\tan^{-1}\left(\f{e^\xi}{\sqrt{1-e^{2\xi}}}\right)\right)}\left[C e^{-\xi(1-\alpha)}\left(1+be^{2\xi}\right)\right],
\ee
where $\zeta$, in this case, is expressed in units of $\sqrt\beta$. We can easily see that our {\it quasiposition wave function} (\ref{quapos}), i.e. the probability amplitude for the particle (Universe) being maximally localized around the position $\zeta$, is nondiverging for all $\zeta$. We stress that the canonical wave function (\ref{solcan}) is diverging at the classical singularity $x=0$.
This makes a first comparison between the GUP and WDW schemes.
\begin{figure}
\centerline{\psfig{file=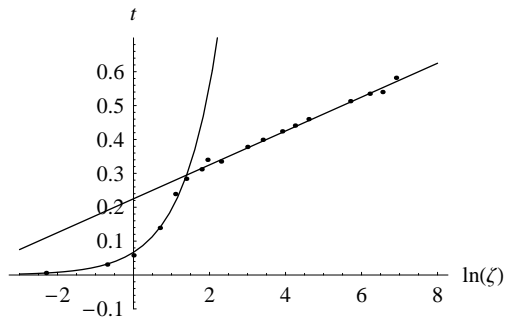,width=7cm}}
\vspace*{8pt}
\caption{The peaks of the probability density $\vert\Psi(\zeta,t)\vert^2$ are plotted as functions of $t$ and $\ln(\zeta)$. The points (resulting from numerical computation) are fitted by a logarithm $0.050\ln(\zeta)+0.225$ for $\zeta\geq4$ and by a power law $0.067\zeta^{1.060}$ for $\zeta\in[0,4]$.} 
\end{figure}

To obtain information on the fate of the Big-Bang singularity in the GUP framework we have to construct and to examine the motion of wave packets
\be\label{wp} 
\Psi(\zeta,t)=\int_0^\infty d\epsilon g(\epsilon)\Psi_\epsilon(\zeta)e^{i\epsilon t},
\ee
where we have defined the dimensionless time $t=\sqrt{24\pi\kappa}\phi$ and $g(\epsilon)$ is a Gaussian distribution peaked at some $\epsilon^\ast\ll1$, which corresponds to be peaked at energy much less then the Plank energy $1/l_P$. The probability density $\vert\Psi(\zeta,t)\vert^2$ to find the Universe around $\zeta\simeq0$ (i.e. around the Planckian region) can be expanded as $\vert\Psi(\zeta,t)\vert^2\simeq\vert A(t)\vert^2+\zeta^2\vert B(t)\vert^2$ and $\vert A(t)\vert^2$ is well approximated by a Lorentzian function packed at $t=0$. Therefore, the probability density to find the Universe in a Planckian volume is peaked around the corresponding classical time and as a matter of fact, it vanishes for $t\rightarrow-\infty$, i.e. where the classical singularity appears. This is the meaning when we claim that the classical cosmological singularity is solved by this model. Of course, the most interesting differences between the WDW and the GUP approaches can be recognized in the wave packets dynamics. In particular, we consider a wave packet initially peaked at late times and let it evolve numerically ``backward in time''. The result of the integration is that the probability density, at different fixed values of $\zeta$, is very well approximated by a Lorentzian function yet. Moreover, the width of this function remains, actually, the same as the states evolves from large $\zeta$ ($10^3$) to $\zeta=0$. The peaks of Lorentzian functions, at different $\zeta$ values, move along the classically expanding trajectory (\ref{clastra}) for values of $\zeta$ larger then $\sim4$. Near the Planckian region, i.e. when $\zeta\in[0,4]$, we observe a modification of the trajectory of the peaks. In fact they follow a power-law up to $\zeta=0$, reached in a finite time interval and they {\it escape} from the classical trajectory toward the classical singularity (see Fig. 1). The peaks of the Lorentzian at fixed time $t$, evolves very slowly remaining close to the Planckian region. Such behavior outlines that the Universe has a stationary approach to the cut-off volume, accordingly to the behavior in Fig. 1. For details see [\refcite{BM07a}]. 

This peculiar behavior of our quantum Universe is different from other approaches to the same problem. In fact, recently, it was shown how the classical Big-Bang is replaced by a Big-Bounce in the framework of Loop Quantum Cosmology (LQC) [\refcite{APS}]. Intuitively, one can expect that the bounce and so the consequently repulsive features of the gravitational field in the Planck regime are consequences of a Planckian cut-off length. But this is not the case. As matter of fact, we can observe from Fig. 1 that there is not a bounce for our quantum Universe. The main differences between the two approaches resides in the quantum modification of the classical trajectory. In fact, in the LQC framework we observe a ``quantum bridge'' between the expanding and contracting Universes; in our approach, contrarily, the probability density of finding the Universe reaches the Planckian region in a stationary way. However a recent computation [\refcite{MS}] suggest how, considering higher order corrections with respect those considered in [\refcite{APS}], a different scenario appears. In particular, the initial cosmological singularity is still preserved and therefore no quantum bounce can takes place.

Let us now extend the above framework to a more general cosmological model, i.e. the Taub one, discussing its quantization in the GUP scheme [\refcite{BM07b}]. The Taub model is a particular case of the Bianchi IX model [\refcite{rev}] (for $\gamma_-=0$), where the latter is described by the line element (in the Misner parametrization) 
\be
ds^2=N^2dt^2-e^{2\alpha}\left(e^{2\gamma}\right)_{ij}\omega^i\otimes\omega^j,
\ee
where $N=N(t)$ is the lapse function, the variable $\alpha=\alpha(t)$ describes the isotropic expansion of the Universe and $\gamma_{ij}=\gamma_{ij}(t)$ is a traceless symmetric matrix which determines the shape change (the anisotropy) {\it via} $\gamma_\pm$. Since the determinant of the 3-metric is given by $h=\det e^{\alpha+\gamma_{ij}}=e^{3\alpha}$, it is easy to recognize that the classical singularity appears for $\alpha\rightarrow-\infty$. As well-known [\refcite{rev}] the dynamics of this Universe, toward the singularity, is described by the motion of a two-dimensional particle (the two physical degree of freedom of the gravitational field) in a dynamically-closed domain. In the Misner picture, such a domain depends on the time-variable $\alpha$, while in the Misner-Chitr$\acute e$ ones, becomes independent on it. The next step is to perform the ADM reduction of the dynamics, i.e. to solve the classical constraint with respect to a given momenta before implementing some quantization algorithm. In particular, in the Poincar$\acute e$ plane the ADM ``constraint'' becomes $-p_\tau\equiv H_{ADM}^{IX}=v\sqrt{p_u^2+p_v^2}$, being $\tau$ and $(u,v)$ the new time and anisotropies variables respectively [\refcite{rev}]. As we said, the Taub model is the Bianchi IX model in the $\gamma_-=0$ case and thus its dynamics is equivalent to the motion of a particle in a one-dimensional closed domain. Such model appears for $u=-1/2$ and therefore its ADM Hamiltonian is 
\be\label{ht}
H_{ADM}^T=p_x\equiv p, \qquad x\in[x_0\equiv\ln(1/2),\infty),
\ee
where $x=\ln v$. The classical singularity now appears for $\tau\rightarrow\infty$. 

Let us now compare the canonical (WDW) and the deformed (GUP) quantum Universes. As in the isotropic setting, the best thing to do in order to understand such differences, is to analyze the motion of suitable wave packets. 

In the canonical case, i.e. for $\beta=0$, the quantum features of the Taub Universe are summarized in the behavior of wave packets reported in Fig. 2.
\begin{figure}
\centerline{\psfig{file=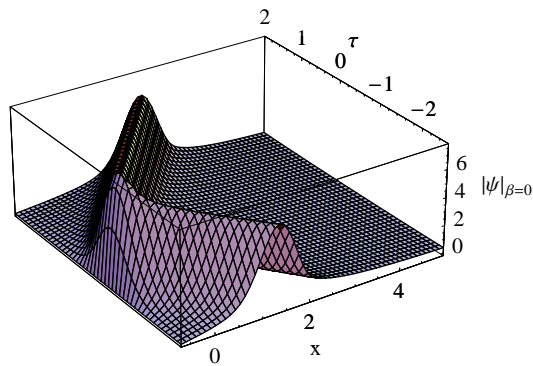,width=7cm}}
\caption{The evolution of the wave packets $\vert\Psi(\tau,x)\vert$ in the WDW framework, $x\in[x_0,5]$.} 
\end{figure}
As we can see from this picture, the wave packets are peaked around the classical trajectories. Therefore the ``incoming'' Universe ($\tau<0$) bounce at the potential wall at $x=x_0$ and then fall toward the classical singularity ($\tau\rightarrow\infty$). This way, the WDW formalism is not able to get light on the necessary quantum resolution of the classical cosmological singularity.

Let us now quantize this model in the GUP framework. Because of the ADM reduction of the dynamics, the variable $\tau$ is regarded as a time coordinate and therefore the conjugate couple ($\tau,p_\tau$) will be treated in a canonical way. This way, we deal with a Schr\"odinger equation $i\p_\tau\Psi(\tau,p)=\hat H_{ADM}^T\Psi(\tau,p)$, where the $\hat H_{ADM}^T$ operator is obtained by using the algebra representation (\ref{rep}). As we said the parameter $\beta$, i.e. the presence of a non-zero minimal uncertainty in the configuration variable, is responsible for the GUP effects on the dynamics and in our case the variable $x=\ln v$, which is related to the Universe anisotropy $\gamma_+$ as
\be\label{anix}
\gamma_+=\f{e^\tau}{\sqrt3v}\left(v^2-\f34\right)=\f{e^{\tau-x}}{\sqrt3}\left(e^{2x}-\f34\right),
\ee
is the configuration variable for the system. Therefore, since the relation $\Delta x_{min}=\sqrt\beta$ appears, we see that the physical interpretation of $\beta$ is to give a non-zero minimal uncertainty in the anisotropy of the Universe. In order to understand the modifications induced by the deformed Heisenberg algebra (\ref{modal}) on the canonical Universe dynamics, we have to analyze different $\beta$-regions. In fact, when the ``deformation'' parameter $\beta$ becomes more and more important, i.e. when we are at some scale which allows us to appreciate the GUP effects, the evolution of the wave packets is different from the canonical case. In particular, for $\beta\sim\mathcal O(1)$ ($k_0=1$), a dominant probability peak close the potential wall appears and the motion of wave packets show a stationary behavior, i.e. they are independent on $\tau$. See Fig. 3.
\begin{figure}
\centerline{\psfig{file=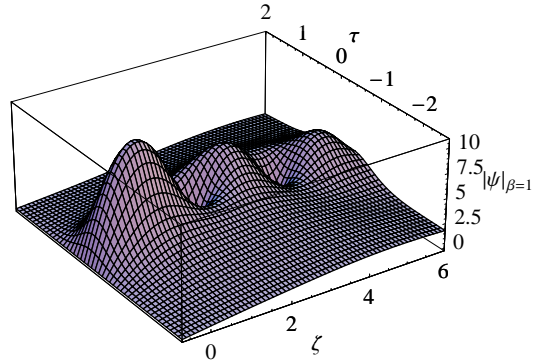,width=7cm}}
\caption{The evolution of the wave packets $\vert\Psi(\tau,\zeta)\vert$ in the GUP framework. For smaller $\beta k_0^2$ the canonical case is recovered. In particular in this plot we choose $k_0=1$.} 
\end{figure}
Therefore two main conclusions can be inferred: (i) The probability amplitude to find the Universe is peaked near the potential wall. In other words, the GUP Taub Universe exhibits a {\it singularity-free behavior}. (ii) The large anisotropy states, i.e. those for $|\gamma_+|\gg1$, are probabilistically suppressed. In fact, from equation (\ref{anix}), the Universe wave function appears to be peaked at values of anisotropy $|\gamma_+|\simeq\mathcal O(10^{-1})$. In this respect, the GUP wave packets {\it predict the establishment of a quantum isotropic Universe} differently from what happens in the WDW theory. For details see [\refcite{BM07b}].

\end{document}